\documentclass[final]{aipproc}
\layoutstyle{8x11single}

\begin{document}

\title 
[Different satellites - different GRB redshift distributions?]
{Different satellites - different GRB redshift distributions?}


\author{Z. Bagoly}{address={ Dept. of Physics of Complex Systems, E\" otv\" os University, H-1117 Budapest, P\'azm\'any P. s. 1/A, Hungary}}
\iftrue
\author{L. G. Bal\'azs}{address={ Konkoly Observatory, H-1525 Budapest, POB 67, Hungary}}
\author{I. Horv\'ath}{address={ Dept. of Physics, Bolyai Military University, H-1581 Budapest, POB 15, Hungary}}
\author{J. Kelemen}{address={ Konkoly Observatory, H-1525 Budapest, POB 67, Hungary}}
\author{A. M\'esz\'aros}{address={ Astronomical Institute of the Charles University, V Hole\v{s}ovi\v{c}k\'ach 2, CZ-180 00 Prague 8, Czech Republic}}
\author{P. Veres}{address={ Dept. of Physics of Complex Systems, E\" otv\" os University, H-1117 Budapest, P\'azm\'any P. s. 1/A, Hungary}}
\author{G. Tusn\'ady}{address={ R\'enyi Institute of Mathematics, Hungarian Academy of Sciences, H-1364 Budapest, POB 127, Hungary}}
\fi

\copyrightyear  {2008}

\begin{abstract}
The measured redshifts of gamma-ray bursts (GRBs), which were first detected by
the Swift satellite, seem to be bigger on average than the redshifts of GRBs
detected by other satellites.  We analyzed the redshift distribution of GRBs
triggered and observed by different satellites (Swift\citep{sak08}, HETE2\citep{van04}, BeppoSax,
Ulyssses). After considering the possible biases {significant difference was
found at the $p=95.70\% $ level in the redshift distributions of GRBs measured
by HETE and the Swift. }
\end{abstract}

\date{\today}

\maketitle



\section{Introduction}

Greiner (\url{http://www.mpe.mpg.de/~jcg}) lists the observations concerning
the afterglows of GRBs, and also selects and lists the confirmed redshifts.  In
\cite{2006A&A...453..797B} we analyzed Swift vs. all non-Swift spacecrafts
redshift data between 01/01/2005-31/01/2006 : five statistical tests show 
$p \ge 99.40$\% significance comparing the redshift distributions for the Swift
and non-Swift samples. It suggested that the redshifts of the Swift sample are
on average larger than that of the non-Swift sample.

Here were extend our work and use GRBs between 28/02/1997 and 03/05/2008 from
Greiner's survey. Since the Ulysses, ASM and XTR trio observed a total of 8
GRBs, therefore we aggregated them into one group (labeled Ulysses).  
The detailed statistics are the following: \\ \ \\
$\begin{array}{l|r|r|l|l}
\hline 
\mbox{Spacecraft} & \mbox{GRB} & \mbox{GRB with $z$} & z_{\mbox{min}}& z_{\mbox{max}}
\\ \hline 
\parbox{8.3cm}{HETE}& 79 & 20 & 0.1606 & 3.372 \cr
\parbox{8.3cm}{SAX} & 57 & 19 & 0.0085 & 3.9 \cr
\parbox{8.3cm}{Swift}& 315 & 103 & 0.0331 & 6.29 \cr
\parbox{8.3cm}{\begin{flushleft} Ulysses group (Ulysses + ASM + XTR) \end{flushleft} } & 64 & 8 & 0.706 & 4.5 \cr
\hline 
\end{array} $

\section{Biases} 

\begin{figure}
\caption{The raw $n(<z)$ cumulative distribution of the different spacecrafts'
GRB observations.}
\includegraphics[height=0.56\textwidth, angle=270]{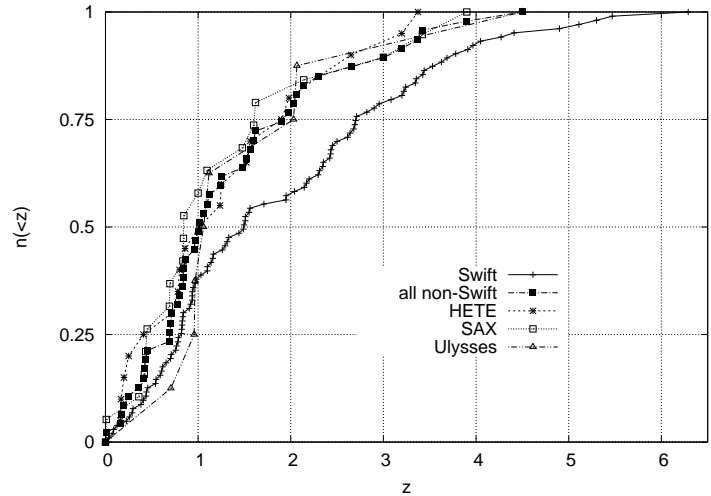}
\end{figure}

To compare the $z$ distributions the Swift and non-Swift samples were
compared using non-parametric rank based tests: the Kolmogorov-Smirnov test and
the median test.  These rank based tests have the clear advantage of being
unaffected by any monotonous transformation in the $z$ values.

The Kolmogorov-Smirnov test compares the maximum difference in the cumulative
distributions of the redshifts in the two samples.  
The median test compares the medians of the Swift and non-Swift samples as
follows: be chosen $N_{\mbox{Swift}}$ objects randomly from the sample of the
non-Swift events ($N_{\mbox{Swift}}$ denotes the number of GRBs in the Swift
sample), and calculate the median. Repeat this e.g. 100000 times, and these
Monte-Carlo simulations give the median distribution for $N_{\mbox{Swift}}$
random GRBs selected from the non-Swift group. Comparing this distribution with
the real Swift median $z$ gives us the significance level for the null
hypothesis that the two medians are equal.

\begin{figure}
\caption{The significance level changes with the length of the dataset}
\includegraphics[height=0.56\textwidth, angle=270]{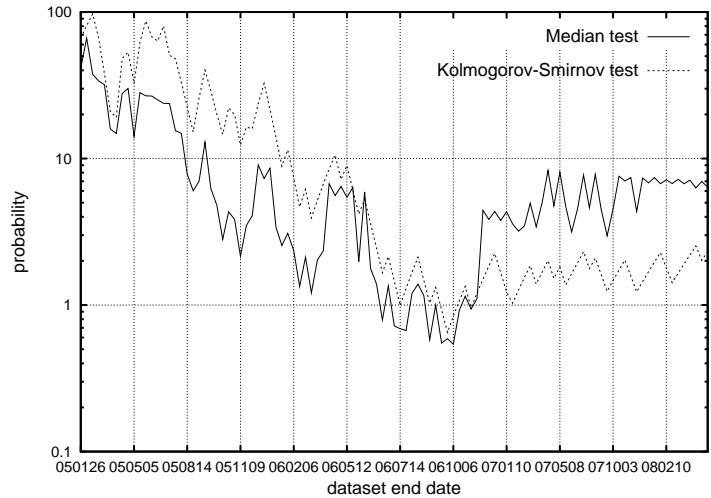}
\label{significance}
\end{figure}

The significance  of the Kolmogorov-Smirnov and the median test changes as new
data arrive continuously from the spacecrafts.  On Fig.\ref{significance}. the
significances' time dependences are shown as a function of the datasets'
end-date. Both values show gradual fall till 10/2006, however after the
probabilities rise - while the length of the datasets grows!  This kind time
dependence indicates some fundamental change in the global observational
strategy.

\begin{figure}
\caption{Swift's GRB density function on the sky. The Voronoi-cell based
density (white: low, black: high) is shown in the equatorial system. The dark
hot spots with high observation probabilies are clearly visible.}
\includegraphics[width=0.46\textwidth, angle=0]{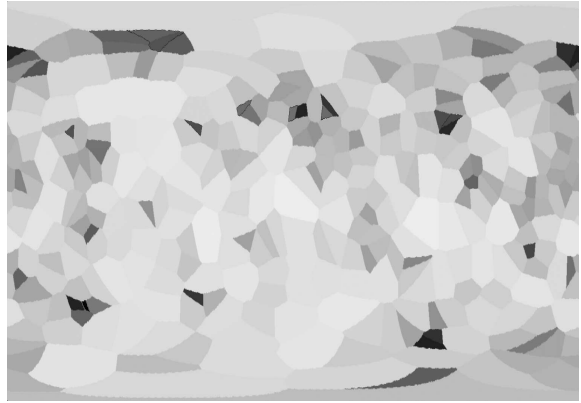}
\label{sky}
\end{figure}

There are definite selections effects from satellite lifespan and sky coverage
E.g. the Swift's X-ray afterglow observations revisit the earlier GRB
directions and create hot spots in the GRB sky distributions on Fig.\ref{sky}. .

The optical follow-up observations' sky coverage is strongly biased biased too,
and it changes from spacecraft to spacecraft. It is due to the different
technical limitations, telescope aviability and the scientific community
interest.

\begin{figure}
\caption{ SAX's redshift-galactic latitude  and 
redshift-declination distribution. 
The galactic disk is clearly visible as a void between $-10 < b < 10$. 
There are some signs of the north/south asymmetry, too.}
\includegraphics[width=0.46\textwidth]{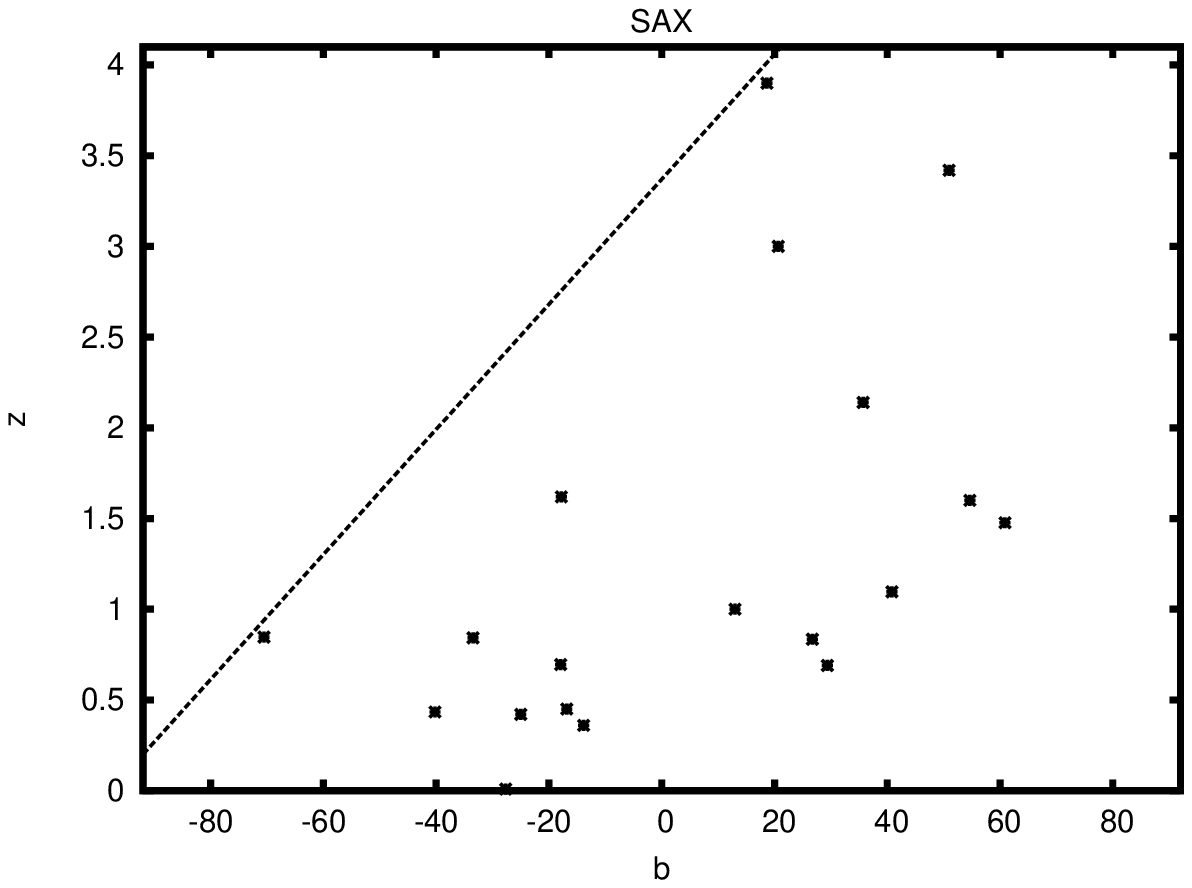}
\includegraphics[width=0.46\textwidth]{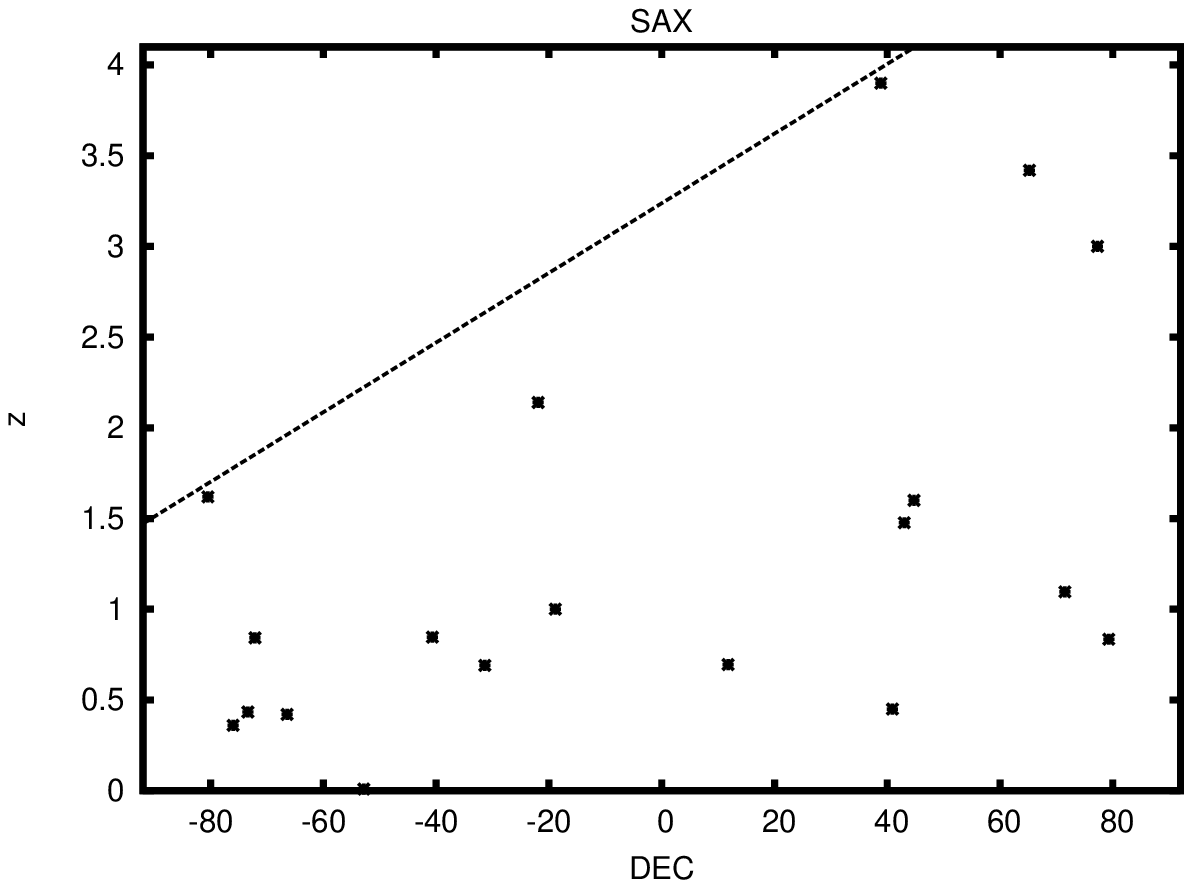}
\label{sax}
\end{figure}

\begin{figure}
\caption{Swift's redshift-declination distribution and a full 3D sky
distribution of the GRBs triggered by the Swift.  The distance from the center
is proportional with $z$, the convex hull of the north and south galactic
hemisphere is also shown. One can observe the void around the galactic plane
and the strong declination dependence, creating a north/south asymmetry.  } 
\includegraphics[width=0.46\textwidth]{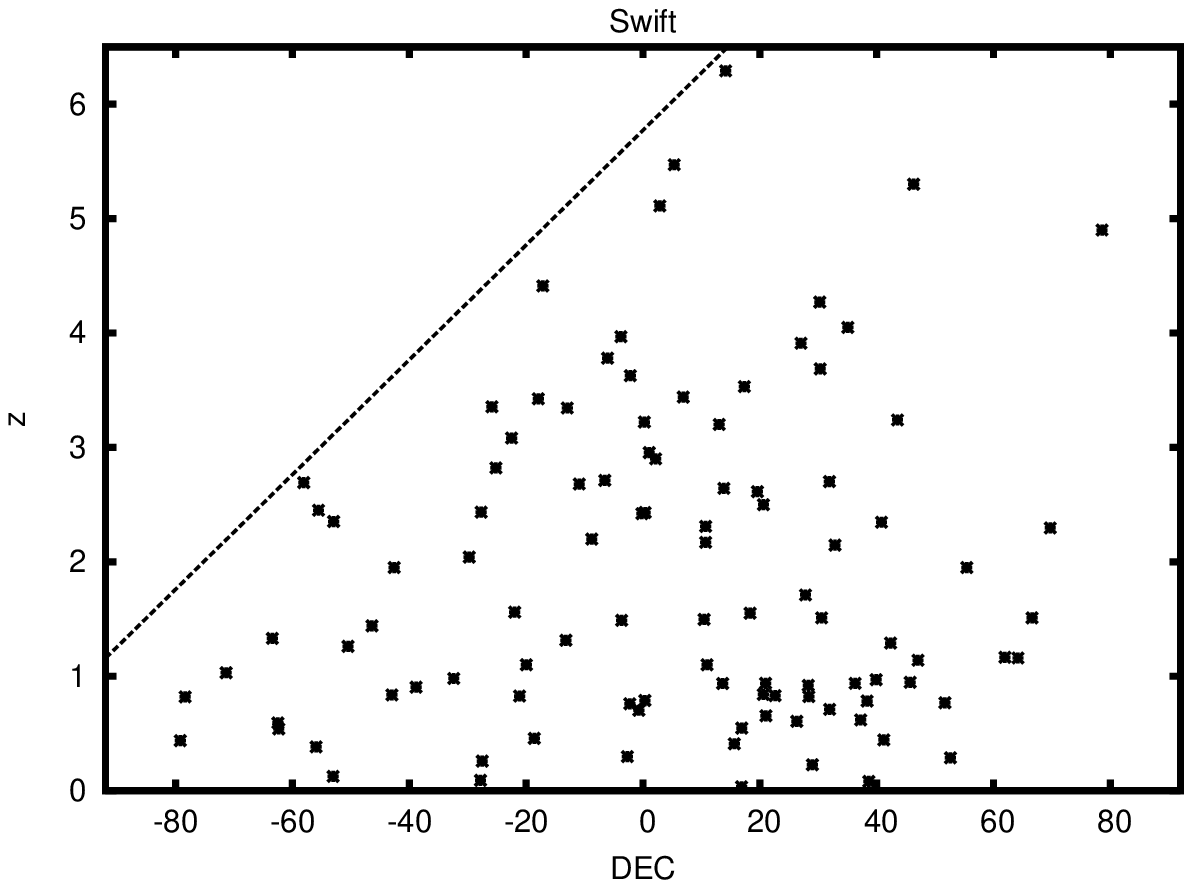}
\includegraphics[width=0.35\textwidth]{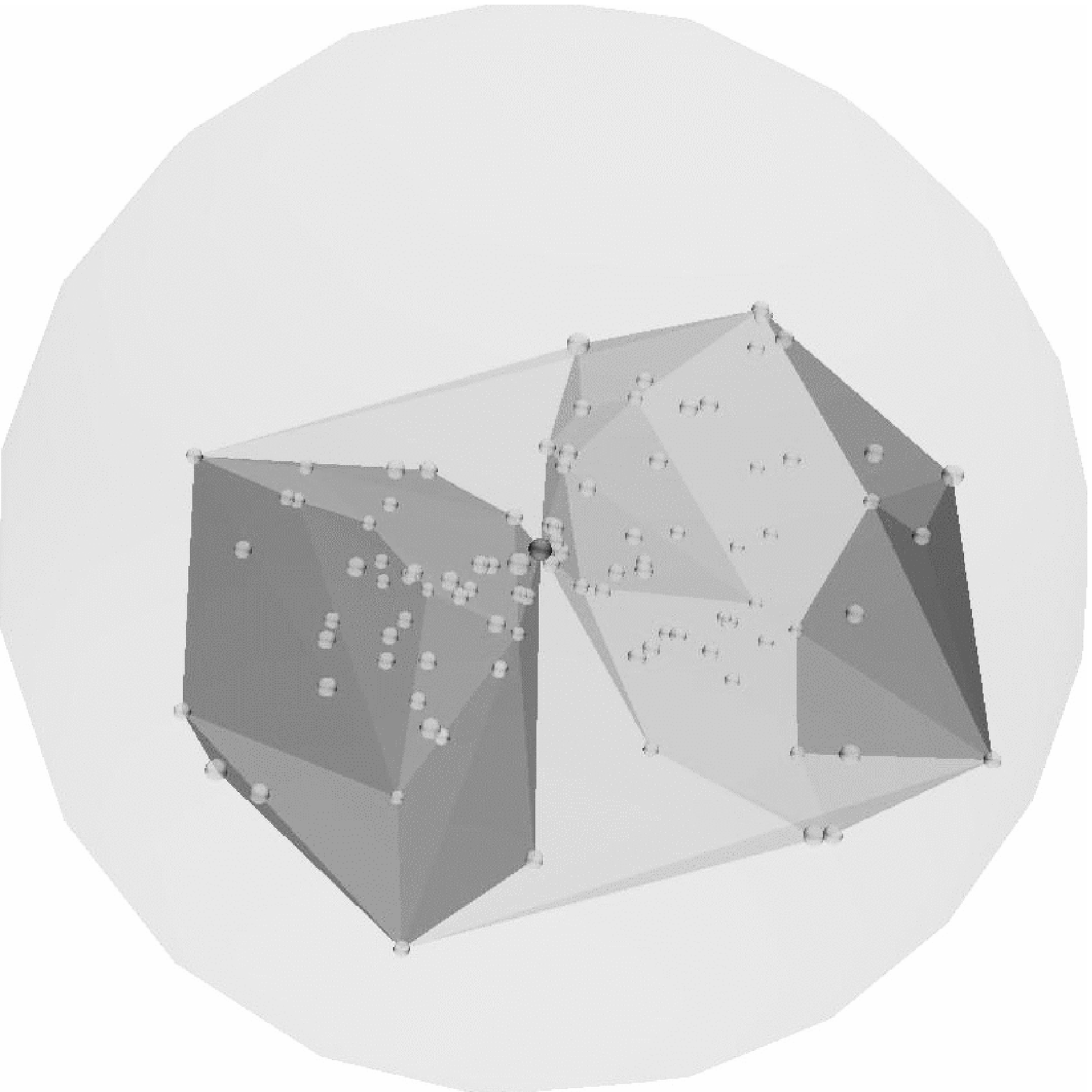}
\label{sw}
\end{figure}

On Figs. \ref{sax}-\ref{sw}. we show the non-isotropic redshift distribution
of the SAX's and Swift's GRBs: both in the galactic and in the equatorial
system there are strong selection effects. The galactic disk is clearly visible
as a void around $-10 < b < 10 $, and the clear cutoff in the redshift at low
declinations shows that the majority of the optical observations were made on
the northern hemisphere.

\section{Reconstruction} 

The observational biases demonstrated in the previous section can be accounted
for - the reconstructions are similar to the magnitude limited quasar sample.  

We used a reconstruction technique based on the Lynden-Bell's C- method
\citep{1971MNRAS.155...95L}, \cite{1987MNRAS.226..273C} to generate weights
from the untruncated part of the data and reconstruct the original
(untruncated) density function. 

Sort burst in ascending order by $z$, and let $\Omega_{i}$ be the solid angle
where all bursts with $z<z_i$ can be detected. In our case $\Omega_{i+1}
\subseteq \Omega_i$, which simplifies the analysis.  We construct the real
$n(<z)$ cumulative density function in the following way: let $N_i=
\sum_{j\in\Omega_i} 1 $, i.e. there are $N_i$ burst within the $\Omega_{i}$
region. Here $n(<z_{i})$ is untruncated, hence $n(<z_{i+1}) = n(<z_{i}) (N_i
+ 1 ) / N_i $.  Starting the sequence with  $n(<z_1)=1$ we can reconstruct the
cumulative density function.

\begin{figure}
\caption{Reconstructed $n(<z)$ distribution of the different spacecrafts' GRB.} 
\includegraphics[height=.56\textwidth, angle=270]{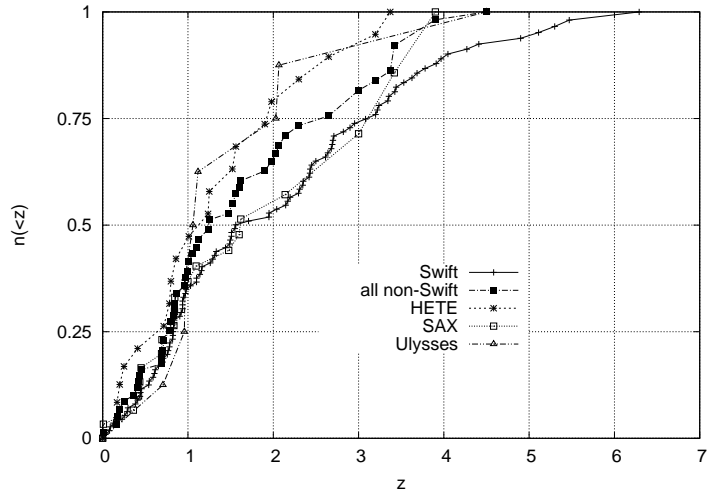} 
\label{reconst}
\end{figure}

For the $\Omega_{z}$ sequences we considered both the $b$ and declination cuts
for each spacecrafts, determined from the real observational data.  
Fig.\ref{reconst}. shows the reconstructed $n(<z)$ cumulative distribution of the different spacecrafts' GRB observations.
Here the KS test gives $p=95.7\%$ for the HETE and Swift distribution. 

\begin{theacknowledgments}
This study was supported by the Hungarian OTKA grant No. T48870, Bolyai
Scholarship (I.H.), by a Research Program MSM0021620860 of the Ministry of
Education of Czech Republic, and by a GAUK grant No. 46307.
\end{theacknowledgments}


\bibliographystyle{aipprocl}
\bibliography{zbagoly}

\end{document}